\documentstyle[preprint,revtex]{aps}
\begin{document}
\font\fortssbx=cmssbx10 scaled \magstep2
\hbox to \hsize{
\hskip.5in \raise.1in\hbox{\fortssbx University of Wisconsin - Madison}
\hfill$\vtop{\hbox{\bf MAD/PH/753}
                \hbox{\bf IFT-P.017/93}
                \hbox{\bf IFUSP-P 1046}
                \hbox{April 1993}}$ }
\vspace{.2in}
\begin{title}
Identifying the Higgs Boson in Electron--Photon Collisions
\end{title}
\author{O.\ J.\ P.\ \'Eboli, M.\ C.\ Gonzalez-Garcia\footnotemark}
\begin{instit}
Instituto de F\'{\i}sica,  Universidade de S\~ao Paulo, \\
Caixa Postal 20516,  CEP 01498-970 S\~ao Paulo, Brazil.
\end{instit}
\footnotetext{\it Permanent Address: Physics Department, University of
Wisconsin, Madison, WI 53706,  USA.}
\author{S.\ F.\ Novaes}
\begin{instit}
Instituto de F\'{\i}sica Te\'orica, Universidade Estadual Paulista \\
Rua Pamplona, 145 -- CEP 01405-900, S\~ao Paulo, Brazil
\end{instit}
\thispagestyle{empty}
\begin{abstract}
We analyze the production and detection of the Higgs boson in the next
generation of linear $e^+e^-$ colliders operating in the $e\gamma$ mode.  In
particular, we study the production mechanism $e + \gamma \rightarrow e \gamma
\gamma \rightarrow e + H$, where one  photon is generated via the laser
backscattering mechanism, while the other is radiated via the usual
bremsstrahlung process.  We show that this is the most important mechanism for
Higgs boson production in a $500$ GeV $e\gamma$ collider for
$M_H\raisebox{-.4ex}{\rlap{$\sim$}} \raisebox{.4ex}{$>$}140$ GeV. We also study
the signals and backgrounds for  detection of the Higgs in the different decay
channels, $b \bar b$, $W^+W^-$, and $ZZ$, and suggest kinematical cuts to
improve the signature of an intermediate mass Higgs boson.
\end{abstract}

\newpage

The most crucial missing element of the Standard Model is the Higgs
boson. Its couplings with all other particles are predicted by the model and
once measured can shed some light on the spontaneous symmetry breakdown
mechanism. In particular, the one-loop  H$\gamma$$\gamma$ coupling receives
contributions from all charged particles that acquire their masses from the
Higgs mechanism, and the study of this coupling can provide us with
fundamental information about the particle mass spectrum. In this work we
analyze the capability of an $e\gamma$ collider to produce and study the
Higgs boson properties, in particular its coupling to photons.

An important feature of the next generation of linear $e^+e^-$ colliders (NLC)
is that they should also be able to operate in the $e\gamma$ or $\gamma\gamma$
modes. The conversion of electrons into  photons  can occur via the laser
backscattering mechanism \cite{las0}. In this case the energy and luminosity of
the photon beam would reach almost the same values of the parent electron beam.
This makes the NLC a very versatile machine which will be able to use energetic
electrons and/or photons as initial states.

In the NLC operating in the $e^+e^-$ mode the most promising mechanisms to
produce the Standard Model Higgs are the Bjorken process, $e^+ + e^-
\rightarrow Z \rightarrow Z + H$, and the vector boson fusion mechanism, $e^+
+ e^- \rightarrow W^+ W^- \rightarrow \nu + \bar{\nu} + H$ \cite{gun}, where
the Higgs boson coupling with $W$'s and $Z$'s could be very well determined.
However, it is  virtually impossible to study the coupling of the Higgs to
$\gamma\gamma$ pairs due to the small branching ratio of Higgs into a photon
pair. Higher order electroweak processes for the Higgs boson production, such
as $e^+ + e^- \rightarrow VVH$ ($V = W, Z$) \cite{bar}, could in principle give
more information on the Higgs boson couplings, but unfortunately their total
cross sections are rather small.

The Standard Model Higgs boson can also be produced at the NLC operating in the
$\gamma\gamma$ mode, as suggested in Ref.~\cite{gunhab}. This mode is
particularly interesting since in principle  it allows a detailed study of the
coupling H$\gamma$$\gamma$. However, as it was pointed out in Ref.~\cite{mea},
resolved  photon processes can impose a severe drawback for the Higgs detection
in the $b\bar{b}$ channel.  Another possibility of Higgs production is
the associated production with the top quark through the process $\gamma\gamma
\rightarrow t\bar{t}H$ \cite{boo}, which is suited for the analyses of the
$t\bar{t}H$ coupling. However, for a 500 GeV collider the total cross section
for this process is below  $0.5$ fb for $M_H > 60$ GeV.

In this letter we concentrate on the possibility of identifying the Higgs boson
with the NLC operating in $e\gamma$ mode.   We stress that this set up provides
us with a rich source of $\gamma \gamma$ interactions  when the hard
backscattered photon interacts with photons radiated by the electrons in the
other beam. This way the Higgs boson can be produced in the process
\begin{equation}
e + \gamma \rightarrow e \gamma \gamma \rightarrow e + H
\label{pro}
\end{equation}
which takes place via the $H\gamma\gamma$ coupling. We show that for a $500$
GeV collider, this is the most important mechanism for the production of a
Higgs with $M_H > 140$ GeV even when compared with the previously suggested
\cite{hag,cheung} associated production $\gamma + e \rightarrow W + \nu_e + H$.
We
study the $b \bar b$, $W^+W^-$, and $ZZ$ signatures of the Higgs and compare
it with the possible backgrounds coming from direct and resolved photon
processes, and from double vector boson production in $e\gamma$ collisions. We
show how convenient kinematical cuts are able to improve in a significant way
the signal over background ratio for the intermediate mass Higgs that decays
mainly through the heavy quark channel.

The scattering of an energetic electron by a soft photon from a laser
allows the transformation of an electron beam into a photon beam whose
spectrum is \cite{las}
\begin{equation}
F_L (x) \equiv \frac{1}{\sigma_c} \frac{d\sigma_c}{dx} =
\frac{1}{D(\xi)} \left[ 1 - x + \frac{1}{1-x} - \frac{4x}{\xi (1-x)} +
\frac{4
x^2}{\xi^2 (1-x)^2}  \right] \; ,
\label{f:l}
\end{equation}
with
\[
D(\xi) = \left(1 - \frac{4}{\xi} - \frac{8}{\xi^2}  \right) \ln (1 + \xi) +
\frac{1}{2} + \frac{8}{\xi} - \frac{1}{2(1 + \xi)^2} \; ,
\]
where $\sigma_c$ is the Compton cross section, and  $\xi \simeq 4
E\omega_0/m^2$, with $m$ and $E$ being the electron mass and  energy. In our
calculation we have chosen the laser energy, $\omega_0$, in order to maximize
the backscattered photon energy without spoiling the luminosity through
$e^+e^-$ pair creation by the interaction between the laser and the
backscattered photon. This can be accomplished by taking $\xi= 2(1+\sqrt{2})
\simeq 4.8$. With this choice, the photon spectrum exhibits a  peak close to
its maximum which occurs at  $x_{max} = \xi/(1+\xi) \simeq 0.83$.

Another source of photons is bremsstrahlung from the electrons in the
other beam. The spectrum of bremsstrahlung photons can be described by
the usual Weizs\"acker--Williams distribution
\begin{equation}
F_{WW}(x)=\frac{\alpha}{2\pi} \frac{1 + (1 - x)^2}{x}
\ln\left(\frac{E^2}{m_e^2}  \right) \; .
\label{WW}
\end{equation}

In the narrow width approximation, the cross section for the process in
Eq.~\ref{pro} can be expressed in terms of the two photon distributions and the
width $\Gamma (H\rightarrow \gamma\gamma)$ as
\begin{equation}
\sigma(e+\gamma \rightarrow e\gamma \gamma \rightarrow e + H)=
\frac{8\pi^2}{s} \frac{\Gamma (H\rightarrow \gamma\gamma)}{M_H}
\int_{\tau_H/x_{max}}^{x_{max}} \frac{dx}{x}~ F_L (x)F_{WW} (\tau_H/x) \; ,
\label{production}
\end{equation}
where $\tau_H=M_H^2/s$ and $\sqrt{s}$ is the center of mass energy of the
original $e^+e^-$ collider. Since we are assuming that all the electron beam
is converted into photons and we have performed  the convolution with the
photon distributions, the corresponding number of events will be given by
$N_{ev}={\cal L}_{e^+e^-}\sigma$, with ${\cal L}_{e^+e^-}$ being the
luminosity of the $e^+e^-$ machine.  The above expression for the cross
section is valid for unpolarized photons. If both the laser backscattered
photon and the electron are polarized and the radiated photon carries part of
the electron polarization, Eq.~\ref{production} should be multiplied by a
factor $(1 +<\lambda_{L} \lambda_{WW}>)$, where $\lambda_{L,WW}$ are the
helicities of the corresponding photons.

In Fig.~\ref{fig:1}.a we plot the cross section  $\sigma(e+\gamma \rightarrow
e\gamma \gamma \rightarrow e + H)$ as a function of the  Higgs mass for
$\sqrt{s}=500$ GeV for two different  top quark masses. For sake of comparison,
we also show the cross section for the process  $e + \gamma \rightarrow W + H +
\nu$ \cite{hag}. As we can see, the  $\gamma\gamma$ intermediate state is a
more efficient way  of producing a Higgs boson with
$M_H\raisebox{-.4ex}{\rlap{$\sim$}} \raisebox{.4ex}{$>$}140$ GeV.  It can lead
to more than 100 Higgs events per year for a Higgs mass up to 320--400 GeV
depending on the top quark mass, assuming a integrated luminosity  ${\cal
L}_{e^+e^-}=100$ fb$^{-1}$.

The cross section for an specific Higgs decay channel is obtained in the narrow
width approximation by the product of the production cross section
(Eq.~\ref{production}) and the corresponding decay branching ratio,
which can be
found elsewhere \cite{gun}. In Fig.~\ref{fig:1}.b we show the cross section for
the processes  $e + \gamma \rightarrow e \gamma \gamma \rightarrow e + H
\rightarrow  e b\bar b (V V^\ast)$, with $V=W^\pm,Z$. The Higgs boson signal
is dominated by different channels depending on its mass, and therefore the
possibility of detecting the Higgs strongly depends on the mass value.

The most promising signal over background ratio is  attained when the Higgs
boson decays into the $ZZ$ channel, which is important for $M_H
\raisebox{-.4ex}{\rlap{$\sim$}} \raisebox{.4ex}{$>$} 180$ GeV. Unlike in the
$\gamma\gamma$ mode of the collider \cite{gunhab} where this channel is free of
irreducible background at tree level, in the $e\gamma$ mode there is a possible
source of background via the process  $e + \gamma \rightarrow e + Z + Z$, with
the final electron going in the beam pipe. In Fig.~\ref{fig:2} we compare the
total  cross section of the Higgs signal measured in fb with the invariant mass
distribution (measured in fb/10~GeV) of the background \cite{bckgZZ}. In the
computation of the  background we assumed an angular size for the beam pipe of
$\theta_e <5 ^\circ$. From Fig.~\ref{fig:2}.a we see that for $500$ GeV
machine, and assuming an invariant mass resolution of the order of $20$ GeV,
the Higgs signal overcomes the background in almost all the $M_H$ range without
need of any further cut. The signal to background ratio in this  channel
becomes better at higher center of mass energies as illustrated in
Fig.~\ref{fig:2}.b.

In the case of the $W^+W^-$ channel the situation becomes worse. The
background from $e + \gamma \rightarrow e + W^+ + W^-$ is large since it
 receives the main contribution from the effective photon process  $e +
\gamma \rightarrow e \gamma \gamma \rightarrow e + W^+ + W^-$,
where, as in the Higgs signal, the electron escapes undetected.  Since
in the $\gamma\gamma$ center of mass frame the signal is  isotropic
while the background is forward-backward peaked \cite{gunhab},   it is
possible to improve the signal to background ratio by imposing a cut  on
the $W$ scattering angle in that system. We chose $|\cos \theta_W|<0.85$. As
seen in Fig.~\ref{fig:3} the background is still one order of magnitude larger
than the signal. Assuming an  invariant mass resolution of $20$ GeV and an
integrated luminosity of $100$ pb$^{-1}$ the signal has a significance of
$3\sigma$ or more only for $M_H\raisebox{-.4ex}{\rlap{$\sim$}}
\raisebox{.4ex}{$<$} 270$ GeV. Although this significance level allow us the
detection of the Higgs in this mode, it makes the observation of the Higgs
properties in this channel a hard task.

For a lighter Higgs boson, $M_H < 150$ GeV, we should rely on the $b\bar{b}$
decay channel. In this case there are large backgrounds coming from the direct
photon process, $\gamma + \gamma \rightarrow b + \bar{b}$, and also from the
once resolved photon ones, $\gamma + \gamma (g) \rightarrow b + \bar{b}$
\cite{mea}, where the photon interacts via its gluonic content. Moreover, there
are also large reducible backgrounds due to the production of charmed quarks,
which can fake $b$-quarks as they possess a production cross section which
exceeds the $b$-pair background by up to one order of magnitude. Again we use
the $S$-wave nature of the Higgs resonance to improve the signal to background
ratio by requiring  that the scattering angle of the $b$'s ($c$'s) with respect
to the beam axis  is large enough $|\cos\theta|<0.85$ in the $b\bar b$ rest
frame.   Imposing this cut in all the following, we compare in
Fig.~\ref{fig:4}.a the total cross section of the Higgs signal $H\rightarrow
b\bar b$, measured in fb, with the invariant mass  distributions (measured in
fb/10~GeV) of the various backgrounds. Since there is a considerable
uncertainty regarding the gluon content of the  photon, we used two  different
sets of gluon distribution functions inside the photon to characterize our lack
of information: the parameterization by Drees and Grassie~\cite{DG} (DG), which
provides for a relatively soft gluon distribution, and the LAC3
parameterization of Ref.~\cite{LAC3}, which gives a considerably harder gluon
distribution.

Figure \ref{fig:4}.b shows the ($y_1+y_2$) distribution in the
laboratory frame for the quarks from signal and  different backgrounds.
In our convention, positive rapidity corresponds to the direction of the
incoming backscattered photon. Both signal and direct backgrounds
populate the positive rapidity region. Since the laser backscattered
photons carry most of the electron beam energy while the bremsstrahlung process
is dominated by soft photons, the jets tend to follow the initial
backscattered photon. Resolved backgrounds, however, present a flat
distribution since they receive two contributions which are
kinematically separated \cite{struct}: in resolved processes where the
bremsstrahlung photon is probed,  the jets follow the initial
backscattered photon and populate the positive  rapidity region, while
in those where laser backscattered photon is resolved the jets  follow
the direction of the bremsstrahlung photon and have negative values of
$(y_1+y_2)$. As a consequence the resolved background can be reduced by
a factor of 50\% approximately by imposing that $(y_1+y_2)>0 $.  We should
remark that this kind of kinematical cut would be worthless in
$\gamma\gamma$ collisions when both photons come from the laser
backscattering mechanism. In this latter case, both the signal and the
resolved background are expected to have approximately the same behavior
over all the rapidity range \cite{mea}.

The final results for $b \bar b$ channel are shown in Fig.~\ref{fig:5},
where an invariant mass resolution of $20$ GeV is assumed for the
backgrounds (for a discussion on this point see Ref.~\cite{mea}).   A
$90\%$ efficiency for both $b$ identification and $c$ rejection  would
lead to a 3--6~$\sigma$ effect for the Higgs boson mass in the range $75
< M_H < 155$ GeV, assuming an integrated luminosity of $100$ fb$^{-1}$
for the NLC with $\sqrt{s}= 500$ GeV. In order to improve the
significance level of the signal and its potential to study the coupling
H$\gamma$$\gamma$, we can consider two possibilities: first we can
reduce the direct background (and increase the signal) by polarizing
both the laser backscattered photon and the electron \cite{gunhab}. Second,
the hadronic calorimeter should be able to give a good coverage close to
the beam pipe in order to detect the jet associated with the remnants of
the resolved photon, allowing to veto the
background coming from the gluonic content of the photon.

In conclusion, we demonstrate that the process $e + \gamma \rightarrow e \gamma
\gamma \rightarrow e + H$  is able to yield a significant number of Higgs
bosons in $e\gamma$ collisions. Furthermore, there is a large window where
even the elusive intermediate mass Higgs boson could be identified when we
take into account the $\gamma\gamma$ fusion mechanism proposed in this paper.
Through this mechanism we should also be able to measure the
$H\gamma\gamma$ coupling, which can give us some hints about the Higgs boson
coupling with all charged particles that acquire mass via
the Higgs mechanism.

\acknowledgments

This work was supported by the University of Wisconsin Research
Committee with funds granted by the Wisconsin Alumni Research
Foundation, by the U.S.\ Department of Energy under Contract
No.~DE-AC02-76ER00881, by the Texas National Research Laboratory
Commission under Grant No.~RGFY9273, by Conselho Nacional de
Desenvolvimento Cient\'{\i}fico e Tecnol\'ogico (CNPq-Brazil),  and by
the National Science Foundation under Contract INT 916182.

M.C. G-G is very grateful to Instituto de F\'{\i}sica Te\'orica de
Universidade Estadual Paulista and Instituto de F\'{\i}sica de Universidade de
S\~ao Paulo for their kind hospitality.

\figure{{\bf (a)} Total cross section at  $\sqrt{s} = 500$ GeV for the process
$e + \gamma \rightarrow e\gamma \gamma \rightarrow e + H$ (solid lines) and
for the process  $e + \gamma \rightarrow W + H + \nu$ (dashed line) versus the
Higgs mass. The lower (upper) curve corresponds to $m_t = 120$ $(200)$
GeV.
\\
{\bf (b)} Total cross section at $\sqrt{s} = 500$ GeV for the process
$e + \gamma \rightarrow e + H  \rightarrow e X$ for different final
states:  $b\bar{b}$ (solid line), $W^+ W^-$ (dashed line), and
$ZZ$ (dotted line) as a function of the Higgs mass. For each type of
line the lower (upper) curve corresponds to $m_t = 120$ $(200)$
GeV.
\label{fig:1}}

\figure{{\bf (a)} Total cross section for the Higgs signal $
H\rightarrow  Z^* Z$  measured in fb (solid line) and invariant mass
distributions (measured in fb/10~GeV) of the background (dashed line)
for  $\sqrt{s} = 500$ GeV. For the background we assume an angular size
for the beam pipe of $\theta_e <5 ^\circ$. The lower (upper) curves
correspond to $m_t = 120$ $(200)$ GeV. \\
{\bf (b)} Same as {\bf (a)} for $\sqrt{s} = 2000$ GeV.
\label{fig:2}}

\figure{{\bf (a)} Total cross section for the Higgs signal $
H\rightarrow  W^* W$  measured in fb for $m_t = 120$ (lower solid line)
and $200$ GeV (upper solid line) and invariant mass distributions
(measured in fb/10~GeV) of the background (dashed line) for  $\sqrt{s} =
500$ GeV. We imposed the cut $|\cos \theta_W| < 0.85$. \\
{\bf (b)} Same as {\bf (a)} for $\sqrt{s} = 2000$ GeV.
\label{fig:3}}

\figure{{\bf (a)} Total cross section for the Higgs signal $
H\rightarrow  b\bar b$  measured in fb (solid line) and invariant mass
distributions (measured in fb/10~GeV) of the various backgrounds for $\sqrt{s}
= 500$ GeV. The dashed lines correspond to the direct photon backgrounds.
Dotted (dash-dotted) lines correspond to resolved photon backgrounds for DG
(LAC3) photon structure functions. In all cases the lower (upper) line
correspond to $b\bar b$ ($c\bar c$) background. We imposed the cut $|\cos
\theta_Q| < 0.85$. \\
{\bf (b)} $(y_1 + y_2)$ distribution in the laboratory frame for the $b\bar{b}$
pairs. The Higgs signal and the different backgrounds are shown for the
invariant mass of 130 GeV for the $b\bar{b}$ pair, using the same cuts and
notation as in {\bf (a)}. \label{fig:4}}

\figure{Cross section for the Higgs signal and backgrounds (same notation as
Fig.~4). For the backgrounds an invariant mass resolution $m_H\pm 10$~GeV is
assumed. In all cases we require $|\cos\theta|_Q<0.85$ in the $Q\bar Q$ rest
frame and $|y_1+y_{2}|>0$ in the laboratory frame.
\label{fig:5}}


\begin{references}
\bibitem{bckgZZ} The cross section for the processes $e + \gamma
\rightarrow V+V+e$, with $V=W$ or $Z$ were obtained through the
numerical evaluation of the amplitudes using a method similar to the one
described in Ref.\ \cite{bar}.
\end{references}
\end{document}